\documentclass[12pt,graphicx,amsmath]{article}
\usepackage{amsmath}
\usepackage{graphicx}
\usepackage{bm}

\usepackage{amssymb}

\title{ Transient Accelerating Scalar Models with Exponential Potential }

\author{W. P.  Cui,  Y.  Zhang,  Z. W. Fu  \\
     \small     Key Laboratory for Researches in  Galaxies  and Cosmology CAS,\\
     \small Department of Astronomy \\
     \small University of Science and Technology of China \\
     \small Hefei, Anhui, 230026,  China }
 \date{}
\begin{document}

\maketitle

\def\vek{\vec{k}}
\renewcommand{\arraystretch}{1.5}
\newcommand{\be}{\begin{equation}}
\newcommand{\ee}{\end{equation}}
\newcommand{\ba}{\begin{eqnarray}}
\newcommand{\ea}{\end{eqnarray}}

\sf

\begin{center}
\Large  Abstract
\end{center}

\begin{quote}

 \sf
\baselineskip=19truept

We study a known class of  scalar dark energy models in which the potential is of
exponential type and the current accelerating era is transient.
We find that, although a decelerating era will return in future,
however, when extrapolating the model back to the earlier stages ($z \gtrsim 4$),
the scalar dark energy becomes dominant over the matter.
So these models do not have the desired tracking behavior,
and the predicted transient period of acceleration
can not be adopted into the standard scenario of Big Bang cosmology.
When couplings between the scalar field and the matter are introduced,
the models still has the problem,
only the time of deceleration return will be varied.
To achieve re-deceleration,
one has to turn to alternative models that are consistent with the standard Big Bang scenario.

\end{quote}

\begin{center}
{\bf 1. Introduction}
\end{center}

Cosmological observations have indicated
that the universe is now in an accelerating expansion
\cite{Riess,Amanullah,CMB,obs,sachswolf,lensing}.
Within the framework of general relativity,
the cause of acceleration can be attributed to
the existence of some dark energy,
which makes up $\sim 70\% $ of the total
cosmic energy in the Universe.
There are a number of possible candidates as the dark energy
driving the accelerating expansion.
The simplest one is the cosmological constant $\Lambda$,
which, however, has the fine-tuning difficulty
and the coincidence problem \cite{Carroll}.
To overcome these difficulties within
the framework of the theory of  general relativity,
dynamic dark energy models have been proposed,
among them, one type is based on some scalar field,
such as
quintessence \cite{quint}, phantom \cite{phantom},
k-essence \cite{k}, Tachyon field \cite{tachyonic},
quintom \cite{quintom}, Chaplygin gas \cite{Chaplygin},
and other type is
dynamic vector models based upon
the Yang-Mills fields \cite{zhang0,Xia,odintsov}.

So far,  there is no observational evidences
whether the current cosmic acceleration is eternal or transient.
Recently, the analysis of combined data of SN Ia+BAO+CMB made
in Ref. \cite{Shafieloo} seems to indicate that the acceleration is slowing down.
Most of dark energy models predict a scenarios that
the acceleration will be eternal.
As is known,
in such an eternally accelerating universe,
there is an event horizon,
and causality exists in a limited spacetime.
From the perspective of string theory,
an asymptotically large space at infinity
is required for the existence of conventional S-matrix,
whose elements are connected with physical observables.
Therefore, S-matrix is ill-defined
in an eternal accelerating universe \cite{Fischler}.
If one wants to save string theory as a theory of physics,
one has to either look for alternatives to the conventional S-matrix,
or to construct cosmological models,
in which the current acceleration is a transient one
and the decelerating expansion will eventually return.

Several dynamic dark energy models have been proposed
to achieve the possibility of cosmic expansion that is accelerating currently,
then will
be decelerating in the future \cite{Townsend,Sahni,Russo,Carvalho,Alcaniz}.
One model was explored in the context of $(4+n)-$dimensional gravity compactified
on an $n$-dimensional time-varying compact manifold \cite{Townsend}.
A class of braneworld models were also shown to admit a transient acceleration
in certain regions of model parameters \cite{Sahni}.
The transient acceleration was also examined
in a scalar cosmological model  in $d$-dimensions
with exponential potentials,
and a general solution was obtained flat Robertson-Walker metric \cite{Russo}.
An extension was made to the case with
a generalized exponential potential of scalar field \cite{Carvalho,Alcaniz}.
Scalar  fields  with  exponential potentials  can occur
generically in  certain theories of particle physics,
such as Kaluza-Klein theory with extra dimensions compactified \cite{Halliwell},
supersymmetry theories \cite{Cremmer},
and higher order gravity \cite{Barrow},
and have been extensively studied
in more general context for dark energy or for inflation as well \cite{Ferreira,Ratra}.

It is interesting to explore
the possibility of transient acceleration
in these general cosmological models.
In particular, we want to check if these models can be extended to earlier epochs
and consistent with the standard Big Bang scenario
as a realistic dark energy model should be.
So far in these scalar models of transient acceleration,
either the cosmic matter component was not taken into consideration,
or the dark energy was assumed to be independent of the matter.
In the time-dependent $\Lambda(t)$ model \cite{Costa}
and the fluid dark energy model \cite{Fabris},
interactions are introduced between dark energy and matter.
Although so far there has not been any observational indication
of a coupling between the dark energy and the matter,
the dark energy as a dynamic scalar field
could have interactions with other components, such as the matter.
We will also study
the models with coupling between scalar dark energy and matter,
and to examine the impacts of the coupling upon the dynamic evolution behavior.

\begin{center}
{\bf 2. The coupling model}
\end{center}

We consider a spatially flat ($k=0$) Robertson-Walker spacetime
with a metric $ds^2=dt^2-a^2(t)d{\bf x}^2$.
The lagrangian of the scalar field
that drives the acceleration is given by
${\cal{L}}=\frac{1}{2}\partial^{\mu}\phi\partial_{\mu}\phi-V(\phi)$.
The energy density and pressure are
$\rho_\phi=\frac{1}{2}\dot{\phi}^2+V(\phi)$, $p_\phi=\frac{1}{2}\dot{\phi}^2-V(\phi)$,
and the equation  of state  $w= p_\phi /\rho_\phi $.
The dynamic expansion of universe is determined by
 the Friedmann equations:
\be \label{friedmann1}
H^2 = \frac{8 \pi G}{3}(\rho_\phi+\rho_m +\rho_r),
\ee
\be \label{friedmann2} \frac{\ddot{a}}{a}
  =-\frac{4 \pi G}{3}(\rho_\phi+3p_\phi+\rho_m +\rho_r +3p_r),
\ee
where $H=\dot a/a$ is the expansion rate
and its present value  $H_0$ is the Hubble constant,
$\rho_m$ and $\rho_r$ are the energy density of matter and radiation, respectively.
The equations of evolution of the three components are given by
\be \label{eq1}
\dot\rho_\phi +3H(\rho_\phi +p_\phi)=Q ,
\ee
\be\label{eq2}
\dot\rho_m +3H \rho_ m=-Q,
\ee
\be\label{eq3}
\dot\rho_r +4H \rho_ r=0,
\ee
where the coupling $Q$ is a generic coupling term
and has a meaning of the energy exchange rate between the dark energy and matter.
The non-coupling model is $Q=0$.
When $Q>0$, the matter transfers energy into the dark energy.
When $Q< 0$, the dark energy transfers energy into the matter.
We do not include the coupling between the dark energy and the radiation,
which will not affect the following conclusion.
Eq.(\ref{friedmann2})  can be derived from
from the set of equations (\ref{friedmann1}), (\ref{eq1}), (\ref{eq2}) and (\ref{eq3}).
It has been shown that, in absence of the matter component
in the non-coupling model \cite{Russo,Carvalho},
for the current accelerating stage of expansion to be transient,
i.e., for the expansion to turn into decelerating again,
the scalar field potential $V(\phi)$ can take on the following form:
\be \label{V}
V(\phi)=\rho_{\phi\,0}[1-\frac{\lambda}{6}(1+\alpha \sqrt{\sigma}\phi)^2)]
\exp{[-\lambda\sqrt{\sigma}(\phi+ \frac{\alpha \sqrt{\sigma}}{2}\phi^2)]},
\ee
 where $\rho_{\phi\,0}$ is a constant energy density,
 $\sigma=8\pi G/\lambda$, and $\alpha$ and $\lambda$
are two dimensionless, positive parameters of the model.
In the context of this paper,
they can take values around
$\alpha$,  $\lambda \sim 1$.
In the limit $\alpha\rightarrow0$
the potential in Eq.(\ref{V})
reduces to an  exponential potential,
$V(\phi) =V_0 \exp{[- \sqrt{8\pi G \lambda }\phi]}$,
a case that was examined in Ref.\cite{Russo}.

For convenience of computation,
it is simpler to rewrite the set of equations,
Eqs.(\ref{friedmann1}), (\ref{eq1}), (\ref{eq2}),
in the following form,
\be \label{h2}
    h^{2}=\frac{U(y)+ x}{1-\frac{1}{2}(\frac{d y}{dN})^{2}},
\ee
\be \label{y}
\frac{d^{2}y}{dN^{2}}-\frac{3}{2}(\frac{dy}{dN})^{3}
+3\frac{dy}{dN}=(\frac{\Gamma}
{\frac{dy}{dN}}+1.5\frac{d y}{dN} x-U^{'}(y))h^{-2},
\ee
\be \label{x}
\frac{d x}{dN}=-\Gamma-3x,
\ee
where
$h\equiv H/H_0$,
$N\equiv\ln a(t)= -\ln (1+z)$,
$y\equiv\sqrt{\frac{8\pi G}{3}}\phi$,
$ x\equiv \rho_m/\rho_c$,
$\Gamma\equiv  Q/H \rho_c $,
and
$U(y)\equiv V(\phi)/\rho_c $.
All of these quantities are dimensionless.
Specifically, we take the model with $\Omega_\phi=0.73$ and $\Omega_m=0.27$
at $z=0$.
The corresponding initial condition at $z=0$
is $x_i=0.27$,
and $\phi_i=0$ and $\dot\phi_i=\sqrt{\lambda \rho_{\phi 0}/3}$.
The dynamical equations is then solved
in the presence of matter component and radiation,
yielding a transient acceleration about $z=0$
followed by the deceleration for the following three typical cases.
Note that the parameter range of $\lambda$ and $\alpha$
for a transient acceleration in our model
differ from those in Refs.\cite{Russo,Carvalho},
which assumed the absence of matter.

\begin{figure}
\centerline{\includegraphics[width=10cm]{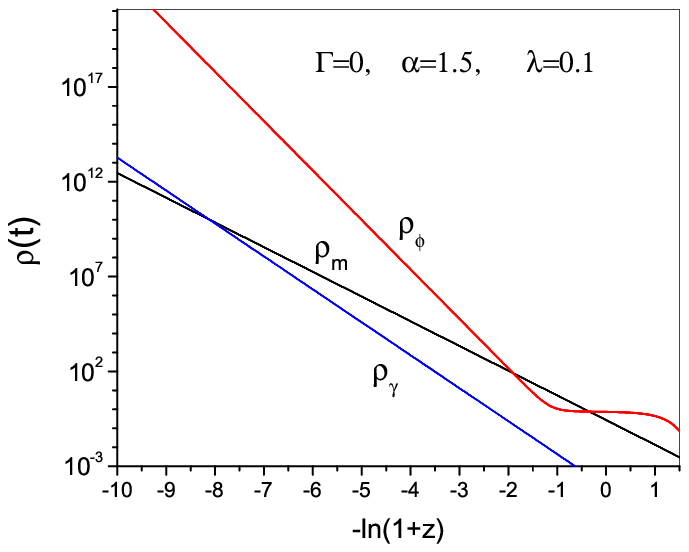}}
\caption{\label{fig01} $\rho(t)$ in non-coupling model
    with $\Gamma=0$.
Notice that $\rho_{\phi}$ dominates $\rho_m$ and $\rho_r$
for $z  \gtrsim 5$,
which is not compatible with the Big-Bang cosmology. }
\end{figure}

\begin{figure}
\centerline{\includegraphics[width=10cm]{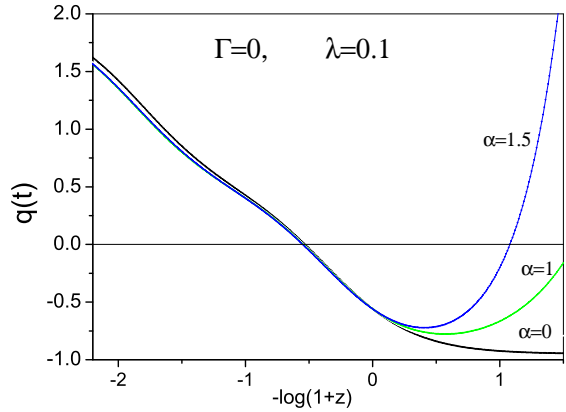}}
\caption{\label{fig3}
A transient acceleration is shown by
the deceleration parameter $q(t)$,
which is negative within $a \sim (0.6, 2.5) $ for $\alpha  =  0.15$.
A greater $\alpha$ yields a shorter duration of acceleration
and an earlier return of deceleration.
}
\end{figure}

For the non-coupling case $\Gamma =0$,
Fig.\ref{fig01} shows the evolution of energy densities,
$\rho_\phi (t)$,  $\rho_m(t)$, and $\rho_r(t)$,
and  Fig.{\ref{fig3}} shows the evolution of
the deceleration parameter $q(z)=- \ddot a/a\dot a ^2$
for various values of the parameter $\alpha$,
larger values of $\alpha>0$
yield a shorter  period of transient acceleration
and an earlier return of deceleration.
The special case with $\alpha=0$ corresponds to
the the exponential potential \cite{Russo},
in which the current acceleration is eternal and deceleration will not return.

\begin{figure}
\centerline{\includegraphics[width=10cm]{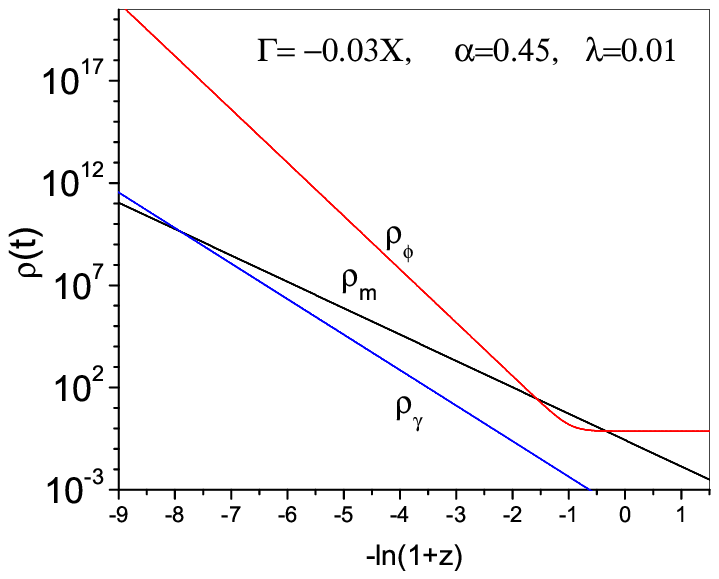}}
\caption{\label{fig6}
The model $\Gamma<0$.
$\rho_{\phi}$ dominates $\rho_m$ and $\rho_r$
for $z  \gtrsim 4$.}
\end{figure}

\begin{figure}
\centerline{\includegraphics[width=10cm]{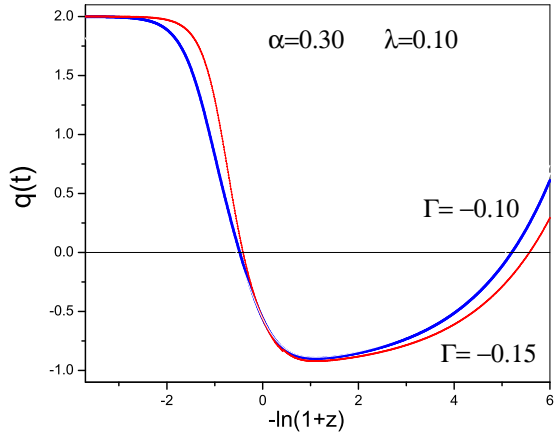}}
\caption{ \label{fig8}
$q(t)$ for various values of  $\Gamma<0$.
$q(t)$ is negative within $a \sim (0.5, 5) $.
}
\end{figure}

Fig. \ref{fig6} show $\rho(t)$,  and Fig. \ref{fig8} show $q(z)$
for the coupling model with the scalar field transfers energy into the matter,
in which
the rate is taken to be proportional to the matter density, $Q \propto -H \rho_m $, i.e.,
$\Gamma \propto -x$.
The dependence upon $\Gamma$ is demonstrated in  Fig. \ref{fig8},
and larger values of  $\Gamma$ yield an earlier return of deceleration.

\begin{figure}
\centerline{\includegraphics[width=10cm]{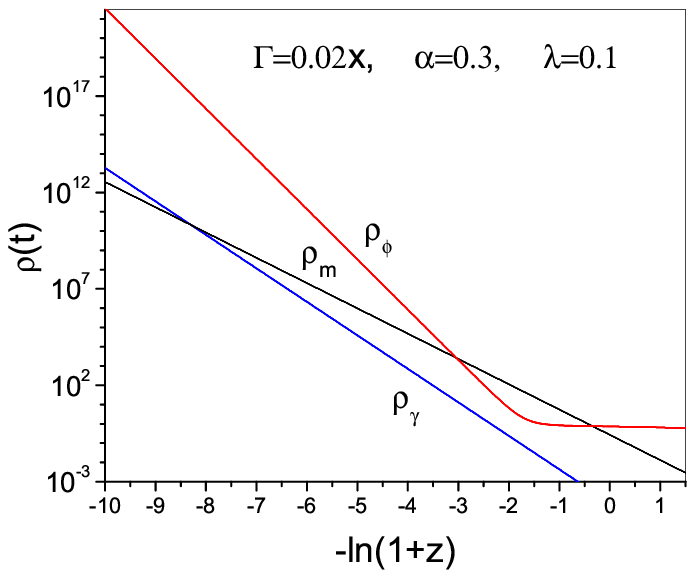}}
\caption{\label{fig9}
The model $\Gamma >0$.
$\rho_\phi(t)$ dominates $\rho_m$ and $\rho_r$
for $z  \gtrsim 20$.}
\end{figure}

\begin{figure}
\centerline{\includegraphics[width=10cm]{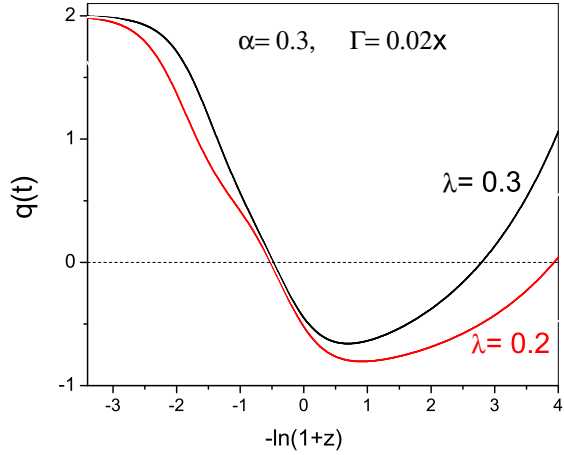}}
\caption{\label{fig11}
The $\Gamma>0$ model for various $\lambda$.
$q(t)$ is negative within $a \sim (0.5, 2.7) $ for $\lambda   =  0.4$.
A greater $\lambda $ yields a shorter duration of acceleration.}
\end{figure}

Figs. \ref{fig9} and \ref{fig11} show the results
for the coupling model in which the matter transfers energy into the scalar field with
$¡¡\Gamma \propto x¡¡$.
The dependence upon the parameter $\lambda $ is demonstrated in  Fig. \ref{fig11},
and larger values of $\lambda$ yield a shorter  period of transient acceleration.

In these plots we extrapolate the model to earlier era.
Figs.\ref{fig01}, \ref{fig6}, and  \ref{fig9} reveal that,
when the models are extrapolated back to the early stages,
$\rho_\phi (t)$ will be dominant over $\rho_m(t)$ and $\rho_r(t)$,
i.e., the $\phi$ field is dominant over the matter,
and has  $w=1$ with $\rho_\phi(t)=p_\phi(t) \propto a^{-6}(t)$,
a feature that has been known \cite{Ferreira}.
The model predicts a scalar field dominated era
for a range of redshift $z > z_a$,
where $z_a$ is some value  $\in( 4 \sim  20)$,
depending upon the sign and value of $\Gamma $ and upon the model parameters.
This kind of dynamic behavior for the early era
deviates drastically from the scenario of the standard cosmology.
Therefore, the class of models with exponential potential
for a transient acceleration
can be pertinent only for rather recent era during the matter domination stage.
They cannot account for the early expansion of the Universe with $z> z_a$.
The standard Big-Bang cosmology has a scenario that
the matter component should be dominant
in the past up to the radiation-and-matter equality
at a redshift $z\sim 3450$ \cite{Spergel},
and the acceleration era starts rather recently around $z\sim 0.5$.
The whole class of models, either $\Gamma=0$, or $\Gamma>0$, or $\Gamma<00$,
have this difficulty.
To be concordant with the Big Bang scenario in this class of models,
one would have to choose a smaller initial value of $\rho_\phi$,
so that $\rho_\phi(t)$ is subdominant to $\rho_m(t)$
during the radiation  or matter dominated era.
Just as investigated in Refs. \cite{Sen,Barreiro},
by constructing  the so-called quintessence $\phi$
with a double exponential potential $V(\phi)$,
one can achieve such a scaling solution followed by an exit
into the accelerating expansion.
However, as we have just shown,
the dynamic evolution of this double exponential  quintessence
does not automatically ensure a proper future deceleration.
If one still wants to achieve a transient acceleration,
one has to use the double exponential  quintessence
for the scaling stage and the exit stage,
and probably employ another new field
for the return of deceleration,   as in Refs. \cite{Russo,Carvalho}.
Still, one would have to give a physical motivation
for such an artificial connection.
It is still premature yet to compare the above models
with the observed data of cosmology,
such as SN Ia \cite{Amanullah}, CMB \cite{wmap7}, and BAO \cite{SDSS percival}.
We mention that
the fluid dark energy model with a transient acceleration \cite{Fabris}
would also face the same problem as the above.

\begin{center}
{\bf 4. Conclusion}
\end{center}

We have explored the possibility of transient acceleration
in a class of scalar dark energy models with exponential potentials
in the presence of the matter and radiation components.
By the detailed examinations for three cases,
the non-coupling with $\Gamma=0 $,
the coupling with $\Gamma<0 $,  and the coupling with $\Gamma >0$,
we find that this class of models with exponential potential
can provide a transient period of acceleration.
However,
extrapolating back to the earlier era
of redshifts
   $z   \gtrsim  5$ $(\Gamma=0 )$,
or $z   \gtrsim  4$  $(\Gamma <0 )$,
or $z  \gtrsim  20$ $(\Gamma >0 )$, respectively,
the scalar energy $\rho_\phi(t)$ will be dominant over
the matter and radiation
and the scaling of behavior is lost too soon.
This would be inconsistent with the standard Big-Bang cosmology,
in which the matter-dominated era extends from $z \sim 3450 $ up to $z\sim  0.5$.
The coupling between the scalar dark energy and the  matter
brings only minor modifications to the dynamic expanding behavior.
A greater value of $\Gamma$ tends to yield a shorter period of acceleration
and an earlier return of deceleration.

Therefore, as they presently stand,
this class of models  with exponential potentials
can be only used at most from more recent past around low redshifts.
In order to have a viable  model of transient acceleration,
one has to either use the exponential scalar potential
to construct more sophisticated models with a  proper dynamic behavior
at high-$z$,
or to seek other models.

\

{ACKNOWLEDGMENT}: W.P Cui's work has been part of
the Undergraduate Research Project of USTC.
Y.Zhang's research work has been supported by the CNSF
No.10773009, SRFDP, and CAS.

\newpage

\end{document}